\documentclass[         %
superscriptaddress,    
nofootinbib,            
onecolumn,             %
floatfix]               
{revtex4-1}               

\def\bra#1{\mathinner{\langle{#1}|}}
\def\ket#1{\mathinner{|{#1}\rangle}}

\usepackage{amssymb}
\usepackage{graphicx}
\usepackage{amsmath}
\usepackage[hypertex]{hyperref}
\usepackage{amsfonts}
\usepackage{natbib}
\usepackage{lipsum}

\usepackage{tabularx}
\usepackage{xcolor}
\usepackage{oubraces}
\usepackage{dsfont}
\usepackage{transparent}
\usepackage{tocloft}
\usepackage{color}
\usepackage{ulem}

\begin{document}
\title{Quantum entanglement shared in hydrogen bonds \\and
its usage as a resource in molecular recognition}
\author{Onur Pusuluk}
\affiliation{Department of Physics, \.{I}stanbul Technical
University, Maslak, \.{I}stanbul, 34469 Turkey}
\author{G\"{o}khan Torun}
\affiliation{Department of Physics, \.{I}stanbul Technical
University, Maslak, \.{I}stanbul, 34469 Turkey} \affiliation{Centre
for the Mathematics and Theoretical Physics of Quantum
Non-Equilibrium Systems, School of Mathematical Sciences, The
University of Nottingham, Nottingham NG7 2RD, United Kingdom}
\author{Cemsinan Deliduman}
\affiliation{Department of Physics, Mimar Sinan Fine Arts
University, \c{S}i\c{s}li, \.{I}stanbul, 34380, Turkey}

\date{\today}

\begin{abstract}

Quantum tunneling events occurring through biochemical bonds are
capable to generate quantum correlations between bonded systems,
which in turn makes the conventional second law of thermodynamics
approach insufficient to investigate these systems. This means that
the utilization of these correlations in their biological functions
could give an evolutionary advantage to biomolecules to an extent
beyond the predictions of molecular biology that are generally based
on the second law in its standard form. To explore this possibility,
we first compare the tunneling assisted quantum entanglement shared
in the ground states of covalent and hydrogen bonds. Only the latter
appears to be useful from a quantum information point of view. Also,
significant amounts of quantum entanglement can be found in the
thermal state of hydrogen bond. Then, we focus on an illustrative
example of ligand binding in which a receptor protein or an enzyme
is restricted to recognize its ligands using the same set of
proton-acceptors and donors residing on its binding site. In
particular, we show that such a biomolecule can discriminate between
$3^n - 1$ agonist ligands if it uses the entanglement shared in $n$
intermolecular hydrogen bonds as a resource in molecular
recognition. Finally, we consider the molecular recognition events
encountered in both the contemporary genetic machinery and its
hypothetical primordial ancestor in pre-DNA world, and discuss
whether there may have been a place for the utilization of quantum
entanglement in the evolutionary history of this system.


\end{abstract}

\maketitle

\section{Introduction}

The conventional thermodynamics is based on the the molecular chaos
hypothesis, which does not allow any correlation between the atoms
and molecules. The famous second law derived within this theory is
fundamental to understanding the biochemistry of cellular processes.
However, there is a tradeoff between correlation and entropy
\cite{1989_PRA_Lloyd_IAndS, 2012_PRL_SagawaAndUeda_CorrInSTD,
2014_PRL_SagawaEtAll_Exp}, due to which the second law, in its
standard form, is insufficient in the presence of nonlocal
correlations \cite{2013_HorodeckiOppenheim_Thermomajorization,
2015_HorodeckiOppenheim_2ndLaws, 2017_Wehner_2ndLaws,
2017_NComms_Winter_UniversalLawsofTD}, e.g., initial quantum
correlations may lead to anomalous heat flows from the cold to the
hot systems \cite{2008_PRE_Partovi, 2010_PRE_JenningsAndRudolph,
2017_arXiv_Lutz}. In this respect, an interesting question to ask is
whether biological molecules can share quantum correlations, which
in turn plays a key role in the cellular functions of these
molecules beyond the conventional thermodynamics.

Quantum correlations can be generated by the electron and proton
tunneling events occurring through the chemical bonds formed between
biological molecules. In particular, a covalent bond is nothing but
the correlated tunneling of two electrons between two atoms.
Consider a single covalent bond formed between two arbitrary atoms
labeled by X and Y. The ground state of this bond can be written as
a coherent quantum superposition \cite{1960_Pauling}:
\begin{eqnarray} \begin{aligned} \label{Eq_State_CB}
|\text{X$-$Y}\rangle = \alpha \underbrace{\big(a
|\text{X}{\color{black} \cdot}\rangle \otimes |{\color{gray}
\cdot}\text{Y}\rangle + b |\text{X}{\color{gray} \cdot}\rangle
\otimes |{\color{black} \cdot}\text{Y}\rangle\big)}_{\equiv
|\text{X}\cdot - \cdot \text{Y}\rangle} + \beta
\underbrace{|(\text{X}{\color{black} \cdot}{\color{gray}
\cdot})^{-}\rangle \otimes |\text{Y}^{+}\rangle}_{\equiv
|(\text{X}{\color{black} \cdot}{\color{black} \cdot})^{-},
\text{Y}^{+}\rangle} + \gamma \, \underbrace{|\text{X}^{+}\rangle
\otimes |({\color{black} \cdot}{\color{gray}
\cdot}\text{Y})^{-}\rangle}_{\equiv |\text{X}^{+}, ({\color{black}
\cdot}{\color{black} \cdot}\text{Y})^{-}\rangle}
\end{aligned} \end{eqnarray}
where all the probability amplitudes depend on distance $r \equiv
d(\text{X}, \text{Y})$ and obeys normalization conditions $|a|^2 +
|b|^2 = 1$ and $|\alpha|^2 + |\beta|^2 + |\gamma|^2 = 1$, the black
and gray dots stand for the electrons participating in the chemical
bond, the former term $|\text{X $\!\cdot\!\! - \!\!\cdot\!$
Y}\rangle$ represents the share of two electrons in the bonding
molecular orbital $\sigma_{\text{X$-$Y}}$ and the subsequent two
terms are responsible for the partial ionic character of the bond.

When the second atom labeled by Y is a hydrogen (H), the last ionic
term in (\ref{Eq_State_CB}) can be ignored because of the weak
electron affinity of this atom \cite{1960_Pauling}. The X$-$H
molecule is then able to form a weak chemical bond with a second
electronegative atom, known as hydrogen bond (H-bond). The atom
which is initially bonded to H is called proton-donor, while the
second atom is called proton-acceptor. Either the electrons of the
proton-acceptor or the proton of the H atom can tunnel back and
forth between the proton-donor and -acceptor.

Consider a H-bond between two arbitrary atoms X$_1$ and X$_2$. This
bond is defined by means of the geometric parameters $r(t) \equiv
d(\text{X}_1, \text{H})$, $R(t) \equiv d(\text{X}_1,\text{X}_2)$,
$\phi(t) \equiv \angle\text{HX}_1 \text{X}_2$. To minimize the
overall energy, the amounts of neutral and ionic contributions to
the ground state of the X$_1-$H bond are initially rearranged in the
presence of X$_2$ as follows:
\begin{align}
|\psi(t_i = 0,R_i = \infty)\rangle &= \Big(\alpha(r)
|\text{X}_1\!\cdot\!\! - \!\!\cdot\!\text{H}\rangle + \beta(r)
|(\text{X}_1{\color{black} \cdot}{\color{black} \cdot})^{-},
\text{H}^{+}\rangle\Big) \otimes
|{\color{black} \cdot}{\color{black} \cdot}\text{X}_2\rangle \label{Eq_State_HB1} \\
&\rightarrow \Big(\alpha^{\prime}(r,R,\phi)
|\text{X}_1{\color{black}\cdot}\!\! - \!\!{\color{black}\cdot}
\text{H}\rangle + \beta^{\prime}(r,R,\phi)
|(\text{X}_1{\color{black}\cdot}{\color{black}\cdot})^{-},
\text{H}^{+}\rangle\Big) \otimes
|{\color{black}\cdot}{\color{black}\cdot}\text{X}_2\rangle \nonumber \\
&\quad= \alpha^{\prime}(r,R,\phi) |\text{X}
_1{\color{black}\cdot}\!\! - \!\!{\color{black}\cdot}\text{H}\rangle
\otimes |{\color{black}\cdot}{\color{black}\cdot}\text{X}_2\rangle +
\beta^{\prime}(r,R,\phi)
|(\text{X}_1{\color{black}\cdot}{\color{black}\cdot})^{-},
\text{H}^{+},
{\color{black}\cdot}{\color{black}\cdot}\text{X}_2\rangle \equiv
|\text{X}_1\!-\!\text{H} \shortmid\shortmid\shortmid
\text{X}_2\rangle , \label{Eq_State_HB2}
\end{align}
where $r(t) > r(0)$ and $|\alpha^\prime(r,R,\phi)|^2 <
|\alpha(r)|^2$ for $t > 0$ in general.

As the changes in the amplitudes described above are only induced by
electrostatic interactions, the state transition from
(\ref{Eq_State_HB1}) to (\ref{Eq_State_HB2}) is completely classical
and is expected to occur for every proton-donor and -acceptor pair.
Besides this, the electron lone pair orbital of the proton-acceptor
and the unoccupied antibonding orbital of the proton-donating bond
($\sigma^{\ast}_{\text{X$_1-$H}}$) may overlap in the ground state
of the whole system, which in turn results in a charge transfer from
the lone pair orbital to $\sigma^{\ast}_{\text{X$_1-$H}}$ in the
form of electron tunneling. This electron delocalization occurring
in the ground state not only weakens and elongates X$_1-$H bond, but
also gives a partially covalent character to the interaction H$\,
\shortmid\shortmid\shortmid \,$X$_2$ \cite{2011_RevCovOFHB}:
\begin{align} |\text{X}_1\!-\!\text{H} \shortmid\shortmid\shortmid
\text{X}_2\rangle &\rightarrow \alpha_{-} |\text{X}
_1{\color{black}\cdot}\!\! - \!\!{\color{black}\cdot}\text{H}\rangle
\otimes |{\color{black}\cdot}{\color{black}\cdot}\text{X}_2\rangle +
\delta_- |(\text{X} _1{\color{black}\cdot}
\raisebox{0.28ex}{\text{$\doteq$}}
{\color{black}\cdot}\text{H})^{-},
({\color{black}\cdot}\text{X}_2)^{+}\rangle +
\beta_{-}|(\text{X}_1{\color{black}\cdot}{\color{black}\cdot})^{-},
\text{H}^{+},
{\color{black}\cdot}{\color{black}\cdot}\text{X}_2\rangle \equiv
|(\text{X}_1\!-\!\text{H})\!-\!\text{X}_2\rangle ,
\label{Eq_State_HB3}
\end{align}
where $|\alpha_-|^2 + |\delta_-|^2 + |\beta_-|^2 = 1$. Here, we omit
the dependence of amplitudes on the geometric parameters for the
sake of simplicity of notation and depict the presence of an
electron in the antibonding orbital $\sigma^{\ast}_{\text{X$_1-$H}}$
by a dot above a second line between X$_1$ and H.

Such orbital interactions are of importance for many H-bonded
systems in biology as their absence may lead to compression of the
X$_1-$H bond rather than its expected elongation
\cite{2002_BlueShiftedHB}. As an example, the covalent contribution
to attractive energy may be comparable to the electrostatic
contribution for interbase H-bonds holding two strands of DNA
together \cite{1999_CovOfHBInDNA, 2000_CovOfHBInDNA,
2006_CovOfHBInDNA}. That is to say, the electron delocalization
through H-bonds may be vital to the stability of Watson-Crick base
pairing. However, it is still too early to reach a consensus on this
matter as the results depend on which quantum chemical model is in
use.

The extent of the covalency of H-bonds is not limited to the lone
pair electrons of the proton-acceptor. The proton which constitutes
the nucleus of the H atom is also likely to be delocalized between
X$_1$ and X$_2$ due to the orbital interactions in the ground state
\cite{1960_Pauling}:
\begin{align} |\text{X}_1\!-\!\text{H}
\shortmid\shortmid\shortmid \text{X}_2\rangle &\rightarrow
\alpha_{+} |\text{X} _1{\color{black}\cdot}\!\! -
\!\!{\color{black}\cdot}\text{H}\rangle \otimes
|{\color{black}\cdot}{\color{black}\cdot}\text{X}_2\rangle +
\delta_+ |(\text{X}
_1{\color{black}\cdot}{\color{black}\cdot})^{-}\rangle \otimes
|(\text{H}{\color{black}\cdot}\!\! -
\!\!{\color{black}\cdot}\text{X}_2)^{+}\rangle + \beta_{+}
|(\text{X}_1{\color{black}\cdot}{\color{black}\cdot})^{-},
\text{H}^{+},
{\color{black}\cdot}{\color{black}\cdot}\text{X}_2\rangle \equiv
|\text{X}_1\!-\!(\text{H})\!-\!\text{X}_2\rangle ,
\label{Eq_State_HB4}
\end{align}
where $|\alpha_+|^2 + |\delta_+|^2 + |\beta_+|^2 = 1$. This is
observed especially in so called low barrier H-bonds (LBHBs) and
short and strong H-bonds (SSHBs). The existence of such bonds in
enzymes and their functional role in biocatalysis are still
controversial \cite{LBHB_1998_MiniRev, LBHB_1998_SSHB, LBHB_2004_No,
LBHB_2006_No, LBHB_2012, LBHB_2013}. Additionally, the nuclei of H
atoms are also likely to tunnel through the interbase H-bonds
\cite{1963_HBPT_Lowdin} and there is another hot debate on the
extent of tunneling based contribution to point mutations that
change genetic information encoded in DNA \cite{2005_Villani,
2006_Villani, 2010_Villani-1, 2010_Villani-2,
2014_BrovaretsAndHovorun-1, 2014_BrovaretsAndHovorun-2,
2015_BrovaretsAndHovorun, 2015_Al-Khalili, 2015_PTInGC}.

In this paper, we take into account the quantum entanglement
generated by the electron and proton tunneling events occurring
through H-bonds in a hypothetical molecular recognition scenario in
which there is a restriction on the number of intermolecular
H-bonds. This enables us to discuss the possible roles of quantum
tunneling in ligand discrimination within the frameworks of quantum
information theory and thermodynamics where the correlations are
routinely regarded as a resource for specific tasks. Before doing
so, we examine the usefulness of quantum entanglement shared between
covalent and H-bonded atoms in the following section.

\section{Quantum entanglement generated through chemical bonds:\\
Covalent vs Hydrogen Bonds}

To provide perceptive insights into the correlations generated by
the motion of electrons or protons in a chemical bond, we will
reduce the complexity by focusing only on the state of particles
participating in the bond in what follows. Furthermore, we will
neglect the spin degrees of freedom to give prominence to the
particle positions. Finally, we will quantify the quantum
correlations found in a generic bipartite state $\rho$ in terms of
its entanglement of formation, a measure defined as
\cite{Wooters-1996}
\begin{equation} \label{Eq_EoF1}
E_F[\rho] = \min \Big( \sum_i p_i E_E\big[| \psi_i \rangle \langle
\psi_i |\big] \Big) ,
\end{equation}
where the minimum is taken over all possible pure state
decompositions that realize $\rho = \sum_i p_i | \psi_i \rangle
\langle \psi_i | $, $E_E$ is the entropy of entanglement that equals
to $E_E[\rho] = S[\rho_{1(2)}] = (S \circ
\mathrm{tr}_{2(1)})[\rho]$, $S$ is the von Neumann entropy defined
as $S[\rho] = - \mathrm{tr}[\rho \, \log_2 \rho]$, and
$\mathrm{tr}_{2(1)}$ is the partial trace over the degrees of
freedom of the second (first) subsystem.

In this respect, one natural choice is to describe the covalent bond
based on a one-qubit representation for the state of an electron
such that when an electron participating in the bond resides in the
atomic orbital of X (Y), it is described by $|0\rangle$
($|1\rangle$). The generic state given in (\ref{Eq_State_CB}) then
becomes:
\begin{eqnarray} \begin{aligned} \label{Eq_Qubit_CB}
|\text{X$-$Y}\rangle = a \alpha |01\rangle_{12} + b \alpha
|10\rangle_{12} + \beta |00\rangle_{12} + \gamma \, |11\rangle_{12}
,
\end{aligned} \end{eqnarray}
where the electrons depicted by black and gray dots in
(\ref{Eq_State_CB}) are labeled respectively by $1$ and $2$.
However, there is a flaw in this description because of the
indistinguishability of the electrons. Consider the pure covalent
bond corresponding to $a |01\rangle_{12} + b |10\rangle_{12}$. This
two-qubit state is entangled for all nonzero values of $a$ and $b$,
i.e., $E_F = - |a|^2 \log_2 |a|^2 - |b|^2 \log_2 |b|^2 > 0$ when
$0<|a|^2<1$ and $0<|b|^2<1$. However, as the two electrons are
identical, it is not possible to discriminate between the states
$|01\rangle_{12}$ and $|10\rangle_{12}$. Hence, although their
coherent superposition originates from the correlated delocalization
of the electrons, it does not possess any quantum entanglement which
is useful from a quantum information and thermodynamics point of
view.

Another problem of this description of the covalent bond is that the
two-qubit state (\ref{Eq_Qubit_CB}) can be either separable or
entangled depending not only on the values of the amplitudes, but
also their signs. For example, the values of $a \alpha = b \alpha =
\beta = \gamma = 0.5$, which correspond to $50$ percent ionic
character in the ground state of the molecule, give zero
entanglement as $|\text{X$-$Y}\rangle = 1/2 \, (|0\rangle_{1} +
|1\rangle_{1}) \otimes (|0\rangle_{2} + |1\rangle_{2})$. When one of
the amplitudes changes its sign, e.g., $\beta$ becomes $- \, 0.5$,
neither the total amount of ionic character of the bond nor the
contribution of each ionic structure to the ground state of the
molecule changes. However, (\ref{Eq_Qubit_CB}) turns out to be an
entangled state in this case as it can not be written in the form of
$(\lambda_1 |0\rangle_{1} + \lambda_2 |1\rangle_{1}) \otimes
(\lambda_3 |0\rangle_{2} + \lambda_4 |1\rangle_{2})$ anymore, where
$\lambda_j$ are proper amplitudes obeying local normalization
conditions. Furthermore, the entanglement between the electrons is
found to be maximal as reduced states of (\ref{Eq_Qubit_CB}) are
maximally mixed for these values, i.e., $\rho_{1(2)} = 1/2
(|0\rangle\langle0| + |1\rangle\langle1|)$ and
$E_F\big[|\text{X$-$Y}\rangle\langle\text{X$-$Y}|\big] = 1$. When a
second sign flip occurs, e.g., $\gamma$ also becomes $- \, 0.5$,
probability of each ionic structure still remains at $0.25$, but the
two electrons get disentangled once again: $|\text{X$-$Y}\rangle =
1/2 \, (|0\rangle_{1} - |1\rangle_{1}) \otimes (- |0\rangle_{2} +
|1\rangle_{2})$. Hence, changes in the amplitudes which do not
affect the partial ionic character of the bond can alter the amount
of quantum entanglement generated through the bond dramatically from
$0$ to $1$ or vice versa according to this representation.

To overcome these two flaws, it is better to move on to a one-qutrit
representation for the state of each atomic orbital such that an
orbital participating in the bond is regarded to exist in state
$|n\rangle$ when it is occupied by $n = \{0, 1, 2\}$ electrons. The
generic state given in (\ref{Eq_State_CB}) then becomes:
\begin{eqnarray} \begin{aligned} \label{Eq_Qutrit_CB}
|\text{X$-$Y}\rangle = \alpha |11\rangle_{XY} + \beta
|20\rangle_{XY} + \gamma \, |02\rangle_{XY} ,
\end{aligned} \end{eqnarray}
where the subscript $X$ ($Y$) stands for the atomic orbital of X
(Y). This two-qutrit state do not possess any entanglement in the
pure covalent case and it is entangled whenever the bond has a
partial ionic character, i.e.,
$E_F\big[|\text{X$-$Y}\rangle\langle\text{X$-$Y}|\big] > 0$ only for
nonzero values of $\beta$ and/or $\gamma$. That is to say, the
electron tunneling is able to create useful entanglement as long as
the ionic character of the covalent bond does not vanish.

Let us start discussing the correlations in H-bonds first in this
context. To do so, we consider the ground state of a classical
H-bond depicted as $|\text{X}_1\!-\!\text{H}
\shortmid\shortmid\shortmid \text{X}_2\rangle$ in
(\ref{Eq_State_HB2}) and rewrite it using the one-qutrit
representation of atomic orbitals:
\begin{align}
|\psi(t_i = 0,R_i = \infty)\rangle &= \Big(\alpha(r)
|11\rangle_{X_1H} + \beta(r) |20\rangle_{X_1H}\Big) \otimes
|2\rangle_{X_2} \nonumber \\
&\rightarrow  \alpha^{\prime}(r,R,\phi) |11\rangle_{X_1H} \otimes
|2\rangle_{X_2} + \beta^{\prime}(r,R,\phi) |20\rangle_{X_1H} \otimes
|2\rangle_{X_2} \equiv |\text{X}_1\!-\!\text{H}
\shortmid\shortmid\shortmid \text{X}_2\rangle .
\label{Eq_Qutrit_cHB}
\end{align}

Since $|\beta^\prime(r,R,\phi)|^2 > |\beta(r)|^2$ for $t > 0$, the
amount of ionic character of the X$_1-$H bond increases due to the
electrostatic interactions with the proton-acceptor X$_2$, so does
the amount of useful entanglement generated by electron tunneling
between X$_1$ and H. Hence, although any fresh entanglement cannot
be created in this type of H-bonds, the entanglement existing in
covalent bonds can be enhanced.

Now, we can move on to the relation between the covalency of H-bonds
and the entanglement in them. To refine this relation, we first
neglect the ionic character of the X$_1-$H bond taking $\beta_-$ in
(\ref{Eq_State_HB3}) to be zero. In this way, we eliminate the
entanglement generated by this single covalent bond. We then extend
the one-qutrit representation to include the molecular orbitals as
well. This enables us to rewrite the ground state
$|(\text{X}_1\!-\!\text{H})\!-\!\text{X}_2\rangle$ in
(\ref{Eq_State_HB3}) simply as:
\begin{align} |\text{X}_1\!-\!\text{H} \shortmid\shortmid\shortmid
\text{X}_2\rangle &= |20\rangle_{\sigma \sigma^{\ast}} \otimes
|2\rangle_{X_2} \nonumber \\
&\quad \rightarrow |2\rangle_{\sigma} \otimes \big(\alpha_{-}
|02\rangle_{\sigma^{\ast} X_2} + \delta_- |11\rangle_{\sigma^{\ast}
X_2} \big) \equiv |(\text{X}_1\!-\!\text{H})\!-\!\text{X}_2\rangle ,
\label{Eq_Qutrit_HB3}
\end{align}
where $\sigma$ and $\sigma^{\ast}$ stand respectively for the
bonding molecular orbital $\sigma_{\text{X$_1-$H}}$ and the
antibonding molecular orbital $\sigma^{\ast}_{\text{X$_1-$H}}$.

The two-qutrit state given in the parenthesis above is entangled
unless $\delta_-$ is equal to zero, which corresponds to vanishing
charge transfer in the form of electron delocalization. This means
that the covalent character of a H-bond implies the formation of
useful entanglement between the H-bonded atoms. Note that the
entanglement generated by electron delocalization between two
covalent-bonded atoms conversely requires a partial ionic character
and is useless otherwise.

Finally, we take into account the quantum entanglement in the
H-bonds involving proton delocalization. To eliminate the
entanglement generated by the single covalent bonds as before, we
neglect their ionic character taking $\beta_+$ in
(\ref{Eq_State_HB4}) to be zero. We then represent the state of the
atom X$_j$ by $|1\rangle$ if it is covalently bonded to the hydrogen
and by $|0\rangle$ otherwise. This one-qubit representation is in
line with the previous treatment of the atomic/molecular orbital
occupancy using a one-qutrit representation and allow us to rewrite
the ground state in (\ref{Eq_State_HB4}) as:
\begin{align} |\text{X}_1\!-\!\text{H}
\shortmid\shortmid\shortmid \text{X}_2\rangle = |10\rangle_{X_1 X_2}
\rightarrow \alpha_{+} |10\rangle_{X_1 X_2} + \delta_+
|01\rangle_{X_1 X_2} \equiv
|\text{X}_1\!-\!(\text{H})\!-\!\text{X}_2\rangle .
\label{Eq_Qubit_HB4}
\end{align}

The amount of the entanglement found in this two-qubit state is
exactly the same as the one found in the three-qutrit state given in
(\ref{Eq_Qutrit_HB3}) when $\delta_+ = \delta_-$, i.e.,
$E_F\big[|\text{X}_1\!-\!(\text{H})\!-\!\text{X}_2\rangle\langle\text{X}_1\!-\!(\text{H})\!-\!\text{X}_2|\big]
=
E_F\big[|(\text{X}_1\!-\!\text{H})\!-\!\text{X}_2\rangle\langle(\text{X}_1\!-\!\text{H})\!-\!\text{X}_2|\big]
= - |\delta_+|^2 \log_2 |\delta_+|^2 - (1 - |\delta_+|^2) \log_2 (1
- |\delta_+|^2)$. The only difference between these two
entanglements is that the former is a correlation directly between
X$_1$ and X$_2$ atoms, while the latter is a correlation actually
between X$_1-$H molecule and X$_2$ atom.

We have neglected the partial ionic character of the X$_1-$H bond to
investigate the quantum entanglement that originates in a H-bonded
system from tunneling of either the electrons of X$_2$ atom or the
proton of the H atom. Although the entanglement generated in this
way was found to be useful from a quantum information and
thermodynamics point of view, its amount obviously decreases with
the raising ionic character of proton-donating covalent bond. Remark
that, contrarily to H-bonds, quantum tunneling of the bonding
electrons was shown to be incapable of creating useful quantum
entanglement in covalent bonds. Instead, covalent bonds were found
to require a particular amount of ionic character to possess a
useful quantum entanglement between their bonded atoms. Hence,
either the tunneling of bonding particles or the ionic character of
the chemical bond has opposite effects on the amount of useful
entanglement shared in covalent and H-bonds.

\section{Environmental effects on correlations: Thermalization \& Decoherence}

Here and in the following, we will describe the many-particle ground
state of a H-bonded system using a unified representation as
follows:
\begin{align} |\epsilon_1\rangle = \alpha_\mp |\psi_1\rangle + \delta_\mp |\psi_2\rangle +
\beta_\mp |\psi_3\rangle , \label{Eq_State_HB5}
\end{align}
where $\mp$ is for discriminating between the cases in which
electron and proton delocalizations occur, $|\psi_1\rangle$ and
$|\psi_3\rangle$ are respectively $|\text{X}
_1{\color{black}\cdot}\! - \!{\color{black}\cdot}\text{H}\rangle
\otimes |{\color{black}\cdot}{\color{black}\cdot}\text{X}_2\rangle$
and $|(\text{X}_1{\color{black}\cdot}{\color{black}\cdot})^{-},
\text{H}^{+},
{\color{black}\cdot}{\color{black}\cdot}\text{X}_2\rangle$ for both
cases, whereas $|\psi_2\rangle$ is either $|(\text{X}
_1{\color{black}\cdot} \raisebox{0.28ex}{\text{$\doteq$}}
{\color{black}\cdot}\text{H})^{-},
({\color{black}\cdot}\text{X}_2)^{+}\rangle$ or $|(\text{X}
_1{\color{black}\cdot}{\color{black}\cdot})^{-}\rangle \otimes
|(\text{H}{\color{black}\cdot}\! -
\!{\color{black}\cdot}\text{X}_2)^{+}\rangle$ depending on which
kind of particle delocalization is under consideration. So far, we
have examined the formation of useful quantum entanglement only on
the basis of this many-particle ground state. However, interaction
with its environment should lead to excitation in this system in
such a way that it reaches a detailed balance between the all energy
eigenstates in the long-time limit as below:
\begin{eqnarray} \label{Eq_State_Th} \begin{aligned}
\rho_{th} &= \sum_{m = 1}^{N} \frac{e^{- \beta
\epsilon_m}}{\mathcal{Z}} | \epsilon_m \rangle\langle \epsilon_m | ,
\end{aligned}
\end{eqnarray}
where $N$ is the total number of energy eigenstates, $\beta$ is the
inverse temperature of the environment that equals to $1/(k_B T)$,
and $\mathcal{Z}$ is the partition function that can be written as
$\sum_m e^{- \beta \epsilon_m}$. All the excited states $|
\epsilon_{m\neq1} \rangle$ should be orthogonal to the ground state
due to the Hermiticity of the many-particle Hamiltonian, which in
turn opens up the possibility that some of the excited states living
inside the subspace $\mathcal{H}_{123}$ spanned by
$\{|\psi_1\rangle, |\psi_2\rangle, |\psi_3\rangle\}$ can be also
entangled to the same extent as the ground state. This may provide a
significant amount of entanglement in the thermal state given in
(\ref{Eq_State_Th}).

To illustrate this argument in detail, suppose that $\alpha_\mp =
\beta_\mp = \delta_\mp = 1/\sqrt{3}$ in (\ref{Eq_State_HB5}) and
there are only two excited states living inside $\mathcal{H}_{123}$
as $| \epsilon_{2} \rangle = (|\psi_1\rangle + |\psi_2\rangle - 2
|\psi_3\rangle)/\sqrt{6}$ and $| \epsilon_{3} \rangle =
(|\psi_1\rangle - |\psi_2\rangle)/\sqrt{2}$. Using the
representation introduced just before (\ref{Eq_Qubit_HB4}), which
gives $|\psi_1\rangle = |10\rangle$, $|\psi_2\rangle = |01\rangle$,
and $|\psi_3\rangle = |00\rangle$, the entanglement of formation is
found to be $0.550048$, $0.187299$, and $1$ respectively for $|
\epsilon_{1} \rangle$, $| \epsilon_{2} \rangle$, and $| \epsilon_{3}
\rangle$. The thermal state is then likely to possess a
non-negligible amount of entanglement for several values of the
Boltzmann factors $\{e^{- \beta \epsilon_m}/\mathcal{Z}\}$, e.g.,
$E_F[\rho_{th}]$ equals to $0.283771$ when
$\frac{1}{\mathcal{Z}}\{e^{- \beta \epsilon_1}, e^{- \beta
\epsilon_2}, e^{- \beta \epsilon_3}\}$ equals to $\{0.7, 0.2,
0.1\}$.

However, the effect of its environment on a quantum system is not
limited to the thermalization, which originates from the exchange of
energy between the system and environment. In addition to this,
information can be also exchanged so that leakage of system's
information into its environment results in decoherence, which
washes out all the entanglement generated inside the system in the
long-time limit. This corresponds to disappearance of the all
non-diagonal elements in the density matrix which describes state of
the system, e.g., $\rho_{th} \rightarrow \rho_d = (22
|00\rangle\langle00| + 19 |01\rangle\langle01| + 19
|10\rangle\langle10|)/60$ for the amplitudes and Boltzmann factors
given in the previous paragraph and this diagonal state has actually
zero entanglement of formation.

\section{Entanglement as a resource in molecular recognition}

Ligand binding typically occurs by weak intermolecular interactions
such as H-bonds, and constitutes one of the major biochemical
functions of proteins. Hence, we will focus on the quantum
entanglement shared between H-bonded molecules in an illustrative
example of ligand binding and investigate the possibility of its
usage as a resource in what follows.

Consider a molecule, a receptor protein or an enzyme, which is
responsible for the recognition of two different ligands. Assume
that this molecule, say that the molecule $A$, binds each of its
ligands through a single H-bond and participates in either
intermolecular H-bond with the same proton-acceptor atom/molecule
X$_1$ residing on its binding site. To challenge $A$ a bit more,
also assume that these two ligands, the molecules $B$ and $C$, share
not only a common molecular shape, but also a similar hydrophobicity
and charge distributions such that even their proton-donating
covalent bonds possess comparable amounts of ionic characters during
the interaction with X$_1$. However, allow a difference in the
extent of electron/proton delocalization events occurring in the two
single H-bonds. Can the molecule $A$ distinguish one ligand from the
other and produces a different agonist response upon binding to each
of them under such circumstances?

The assumptions given above make binding affinities very close to
each other in both cases, which in turn puts a strain on the
distinguishability of the ligands unless $A$ is somehow capable of
identifying the entanglements shared in the two intermolecular
H-bonds. To explore this capability within the framework of quantum
information theory, we will take into account the following states
\begin{align} |\text{X}_B\!-\!(\text{H})\!-\!\text{X}_1\rangle
&= \frac{1}{\sqrt{3}} ( |\psi_1\rangle + |\psi_2\rangle +
|\psi_3\rangle ) , \label{Eq_State_HB_AB} \\
|\text{X}_C\!-\!(\text{H})\!-\!\text{X}_1\rangle &=
\frac{1}{\sqrt{6}} ( |\psi_1\rangle - 2 |\psi_2\rangle +
|\psi_3\rangle ) , \label{Eq_State_HB_AC}
\end{align}
by which it is provided that i) ionic contribution to the
proton-donating covalent bond is exactly the same as neutral
contribution in each intermolecular H-bond, and ii) rate of back and
forth proton tunneling between the proton-acceptor and -donor is
doubled in the second H-bond when compared to the first H-bond. Note
that in spite of this difference in the tunneling rates, the amounts
of entanglement shared in these two bonds are equal, i.e.,
$E_F\big[|\text{X}_B\!-\!(\text{H})\!-\!\text{X}_1\rangle\langle\text{X}_B\!-\!(\text{H})\!-\!\text{X}_1|\big]
=
E_F\big[|\text{X}_C\!-\!(\text{H})\!-\!\text{X}_1\rangle\langle\text{X}_C\!-\!(\text{H})\!-\!\text{X}_1|\big]
= 0.550048$.

To perform its biochemical function successfully, the molecule $A$
is expected to undergo a different conformational transition
depending on to which ligand it binds. To do so, it should first
single the states (\ref{Eq_State_HB_AB}) and (\ref{Eq_State_HB_AC})
out from the other states living inside $\mathcal{H}_{123}$ and then
discriminate between them. As these two states are orthogonal to
each other, they are physically distinguishable. However, $A$ is
able to access the information in them locally, i.e., the reduced
state of X$_1$ equals to
$(2|0\rangle\langle0|+|0\rangle\langle1|+|1\rangle\langle0|+|1\rangle\langle1|)/3$
in the joint system X$_B$X$_1$, while it equals to
$(5|0\rangle\langle0|+|0\rangle\langle1|+|1\rangle\langle0|+|1\rangle\langle1|)/6$
in the joint system X$_C$X$_1$. These two density matrices are not
orthogonal to each other, and so they are not perfectly
distinguishable. Moreover, there are various states living inside
$\mathcal{H}_{123}$ which have the same reduced states with either
(\ref{Eq_State_HB_AB}) or (\ref{Eq_State_HB_AC}) and some of them
are even not entangled.

Despite the fact that the molecule $A$ cannot recognize the
entanglement found in a H-bond formed with another molecule, it can
teleport this entanglement into one of its intramolecular H-bonds by
means of local operations on each molecule and classical
correlations between them. In this respect, imagine that there is a
proton-donating X$_2-$H bond in the close neighborhood of X$_1$
within the binding site of $A$, but the ground state conformation of
$A$ does not allow any orbital interaction between X$_1$ and X$_2$,
e.g., joint state of these atoms/molecules in the ground state
conformation $\chi_1$ equals to
\begin{align} |\text{X}_2\!-\!\text{H}\!\shortmid\shortmid\shortmid\!\text{X}_1\rangle
&= |\epsilon_1\rangle_{\text{X}_2\text{X}_1} \equiv
\frac{1}{\sqrt{2}} (|\psi_1\rangle - |\psi_3\rangle ) .
\label{Eq_State_HB_A1}
\end{align}

On the contrary, assume that the two lowest excited state
conformations of  $A$, $\chi_2$ and $\chi_3$, switch on the orbital
interaction between X$_1$ and X$_2$ such that they stabilize the
intramolecular H-bond in the states which are exactly matching up
with (\ref{Eq_State_HB_AB}) and (\ref{Eq_State_HB_AC}) respectively
as below:
\begin{align} |\text{X}_2\!-\!\text{H}\!\shortmid\shortmid\shortmid\!\text{X}_1\rangle
&\overset{\chi_2}{\longrightarrow}
|\epsilon_2\rangle_{\text{X}_2\text{X}_1} \equiv \frac{1}{\sqrt{3}}
( |\psi_1\rangle + |\psi_2\rangle +
|\psi_3\rangle ) , \label{Eq_State_HB_A2} \\
|\text{X}_2\!-\!\text{H}\!\shortmid\shortmid\shortmid\!\text{X}_1\rangle
&\overset{\chi_3}{\longrightarrow}
|\epsilon_3\rangle_{\text{X}_2\text{X}_1} \equiv \frac{1}{\sqrt{6}}
( |\psi_1\rangle - 2 |\psi_2\rangle + |\psi_3\rangle ) .
\label{Eq_State_HB_A3}
\end{align}

In a sense, the conformation of $A$ is regarded to continuously
monitor the state of joint X$_2$X$_1$ system such that when $A$ is
in a particular conformation $\chi_j$, these H-bonded atoms are
enforced to be in the state $|\epsilon_j\rangle$. This kind of
conditional dynamics can be described by a unitary operation defined
as:
\begin{align}
U_A \big(|\psi(t_i)\rangle_{\text{X}_2\text{X}_1} \otimes
|\phi(t_i)\rangle_{\chi} \big) = \sum_{j=1}^3 \big(M_j
|\psi(t_i)\rangle_{\text{X}_2\text{X}_1}\big) \otimes |\chi_j\rangle
, \label{Eq_State_UA}
\end{align}
where $|\phi(t_i)\rangle_{\chi}$ is the initial conformational state
of $A$, $M_j = |\epsilon_j\rangle
\langle\epsilon_j|_{\text{X}_2\text{X}_1}$ and
$\langle\chi_j|\chi_{j^\prime}\rangle = \delta_{jj^\prime}$. Note
that reduced dynamics of the joint X$_2$X$_1$ system is nothing but
a complete measurement since $\sum_j M_j^\dagger M_j$ equals to
$\mathds{I}_{123}$, the identity operator acting on
$\mathcal{H}_{123}$.

The biochemical transformations starting upon ligand binding in the
molecule $A$ can then be described as a kind of entanglement
swapping \cite{1993_PRL_EntSwap, 1998_PRA_EntSwap}, a key element of
various schemes in quantum information and quantum communication
based on the extension of quantum teleportation
\cite{1993_PRL_qTeleport}, as follows. After forming an
intermolecular H-bond with a molecule $N$, $A$ tries to swap the
partner of X$_1$ in this bond with its atom X$_2$ using a
measurement on the partner atom X$_N$ followed by some operations on
the joint X$_2$X$_1$ system that are classically correlated with the
measurement outcome (see Appendix for the details). In the case of
binding to the ligand $B$ ($C$), perfect swapping of the entangled
partner of X$_1$ brings the joint state of X$_2$ and X$_1$ into the
eigenstate $|\epsilon_2\rangle$ ($|\epsilon_3\rangle$) at time
$t_i$, which demands an accompanying conformational transition from
$\chi_1$ to $\chi_2$ ($\chi_3$) according to (\ref{Eq_State_UA}).
Conversely, when the bonded molecule is neither $B$ nor $C$, state
of the intermolecular H-bond is not likely to match up with one of
the eigenstates of joint X$_2$X$_1$ system. As a matter of fact, it
is expected to be a quantum coherent superposition of them in
general. So the joint X$_2$X$_1$ state at the end of entanglement
swapping process at time $t_i$ is expected to be a quantum coherent
superposition in the basis of $\{|\epsilon_1\rangle,
|\epsilon_2\rangle, |\epsilon_3\rangle\}$, e.g.,
$|\psi(t_i)\rangle_{\text{X}_2\text{X}_1} = \sum_j \lambda_j
|\epsilon_j\rangle$. Its closed system dynamics given in
(\ref{Eq_State_UA}) then attempts to drive the molecule $A$ to the
state $\sum_j \lambda_j |\epsilon_j\rangle \otimes |\chi_j\rangle$,
which is unlikely to occur because of the huge number of atoms in
$A$ that are subjected to decoherence inside the cellular
environment. Hence, any attempt at generating a physiological
response in the form of conformational change should be unsuccessful
in the case of binding to a molecule different than $B$ or $C$. As
an example, imagine an antagonist ligand $D$, which forms the
following H-bond with the molecule $A$:
\begin{align}
|\text{X}_D\!-\!(\text{H})\!-\!\text{X}_1\rangle &=
\frac{1}{\sqrt{6}} ( |\psi_1\rangle + 2 |\psi_2\rangle +
|\psi_3\rangle ) . \label{Eq_State_HB_AD}
\end{align}

As it possesses not only an identical ionic character, but also
completely the same tunneling probability with the molecule $C$, the
molecule $D$ mimics $C$ unless $A$ benefits from the entanglement
swapping to identify the bonded molecule. In this case, the molecule
$A$ is expected to reach the state $(2\sqrt{2}
|\epsilon_2\rangle_{\text{X}_2\text{X}_1} \otimes |\chi_2\rangle -
|\epsilon_3\rangle_{\text{X}_2\text{X}_1} \otimes |\chi_3\rangle)/3$
at the end of a molecular recognition process based on the
entanglement swapping. This corresponds to a quantum coherent
superposition of the whole molecule $A$, which is highly improbable
under the influence of quantum decoherence arising from the
interaction with the cellular environment. Hence, $D$ fails in its
attempt to trigger a conformational change in the molecule $A$, and
the molecular recognition breaks down.

We have taken into account the proton delocalization in
ligand-protein H-bonds so far, but note that our conclusions on the
possibility of the usage of entanglement as a resource in ligand
recognition also holds when we consider the electron delocalization
in these bonds.

\section{Is there a room for quantum measurement in biology?}
\label{Sec_Measurments}

The molecule $A$ appears to be able to make quantum measurements
either on itself or on its ligands in different steps of the
molecular recognition scenario described above. However, there is a
lack of consensus on how the quantum measurement affects the state
of a simple physical system, which is reflected by the wide range of
interpretations of quantum mechanics from Copenhagen interpretation
to many-worlds interpretation. In this respect, we should be careful
about what we mean by a quantum measurement within the framework of
molecular biology.

The reduced dynamics of the joint X$_2$X$_1$ system during the
unitary evolution $U_A$ given in (\ref{Eq_State_UA}) is
mathematically equivalent to a joint measurement in the basis of
$\{|\epsilon_j\rangle\}$ as mentioned before. This reduced dynamics
actually emerges from $U_A$ as a consequence of the elimination of
conformational degrees of freedom of the molecule. That is, the
state of H-bond found between the atoms X$_2$ and X$_1$ is measured,
in a sense, by means of the state of conformation in the course of
closed system dynamics of the molecule. Hence, this measurement can
be regarded as an influence of the conformational dynamics on the
H-bond under consideration.

Besides this, initialization of the teleportation of intermolecular
entanglement shared in X$_N-$(H)$-$X$_1$ bond to the intramolecular
X$_2-$H$\shortmid\shortmid\shortmid$X$_1$ bond requires a
measurement on X$_N$ in the basis of $\{\lambda_1 |1\rangle +
\lambda_2 |0\rangle, \lambda_2^* |1\rangle - \lambda_1^*
|0\rangle\}$ (see Appendix for the details). Note that the
probability amplitudes $\lambda_1$ and $\lambda_2$ obey the
normalization condition $|\lambda_1|^2+|\lambda_2|^2 = 1$, while the
computational basis states $|1\rangle$ and $|0\rangle$ stand for
respectively
$|\text{X}_N{\color{black}\cdot}\!-\!{\color{black}\cdot}\text{H}\rangle$
and $|(\text{X}_N{\color{black}\cdot}{\color{black}\cdot})^-,
\text{H}^+\rangle$. On that account, this measurement operationally
enforces the X$_N-$H covalent bond to display a particular amount of
ionic character at either $(1 - |\lambda_1|^2) \times 100$ or
$|\lambda_1|^2 \times 100$ percentage. It may be physically
impossible for a number of proton-donors to display such an ionic
character, which in turn halts the entanglement swapping at the very
beginning for some ligands. Hence, the initial measurement on X$_N$
may serve as a first-line discrimination mechanism for the molecule
$A$ to single out the ligands having the right amounts of ionic
character from the others.

On the other hand, entanglement swapping between  X$_N-$(H)$-$X$_1$
and X$_2-$H$\shortmid\shortmid\shortmid$X$_1$ bonds does not need
any further quantum operation on the ligand molecule. We can then
reasonably assume that the basis states of the measurement under
consideration coincide with the states of molecules $B$ and $C$ in
the absence of any interaction with $A$ so that the measurement
projects the two agonist ligands into their initial states.
Actually, reorganization of the binding site of $A$ should be
capable to provide such a charge stabilization on X$_N$. Hence, this
measurement can be regarded as a backwash effect of the
conformational dynamics of the molecule $A$ on the proton-donating
X$_N-$H covalent bond.

In this way, we argue that the conformational change of a receptor
protein or an enzyme may affect the intramolecular chemical bonds
(either in itself or in its ligands) to such an extent that it can
impose a strict control on the state of bonded atoms which
mathematically corresponds to a quantum measurement. However, by
doing so, we do not interpret any biological molecule or its
environment as an observer or measurement device. As a matter of
fact, each measurement performed on the quantum state of covalent or
H-bonded atoms in our molecular recognition scenario can be
characterized as an emergent property of the complex chemical
interactions occurring in vivo, which are outside the scope of the
present paper.

\section{Optimum number of hydrogen bonds in a molecular search}

We have proposed a molecular recognition scenario in which a
receptor protein or an enzyme can, at least in principle, harvest
the entanglement shared in an intermolecular H-bond to generate a
physiological response in the form of conformational change. The
number of agonist ligands discriminated by a single H-bond was $2$
in this scenario. To discuss the efficiency of the recognition
process under consideration, we will extend it to the case of more
than two agonist ligands discriminated by multiple H-bonds.

In this respect, we assume that the molecule $A$ is responsible for
the recognition of $N$ different ligands using the same set of
proton-acceptors and donors $\{\text{X}^k_1\}$ residing on its
binding site. We let $n \geq 2$ of the intermolecular H-bonds formed
between $(\text{X}^k_1, \text{X}^k_N)$ pairs to have partially
covalent character, but do not allow any correlation between them.
If $A$ is capable to teleport the entanglement shared in each of
these $n$ intermolecular H-bonds to one of its distinct
intramolecular H-bonds, it can then discriminate between $N = 3^n -
1$ ligands when the unitary evolution $U_A$ given in
(\ref{Eq_State_UA}) is extended into:
\begin{align}
U_A \big(\bigotimes_{k=1}^{n}
|\psi(t_i)\rangle_{\text{X}^k_2\text{X}^k_1} \otimes
|\phi(t_i)\rangle_{\chi} \big) = \sum_{j=1}^3 \bigotimes_{k=1}^{n}
\big(M^k_j |\psi(t_i)\rangle_{\text{X}^k_2\text{X}^k_1}\big) \otimes
|\chi_{j+3(k-1)}\rangle , \label{Eq_State_UA_Ext}
\end{align}
where $M^k_j = |\epsilon^k_j\rangle
\langle\epsilon^k_j|_{\text{X}^k_2\text{X}^k_1}$ and
$\langle\chi_{j+3(k-1)}|\chi_{j^\prime+3(k^\prime-1)}\rangle =
\delta_{jj^\prime} \delta_{kk^\prime}$. That is to say that,
formation of $2$ intermolecular H-bonds involving proton or electron
delocalization is sufficient to single out $8$ agonist ligands from
an infinite number of antagonist ligands. Addition of one more
partially covalent H-bond increases the maximum number of
discernible ligands from $8$ to $26$. This corresponds to an
enormously high efficiency. Let us compare it to the efficiency of
genetic machinery, which is expected to operate quite accurate and
very fast molecular recognition events.

First and foremost, DNA and RNA molecules that carry the genetic
information inside a cell are made up from only $4$ of the tens of
nucleobases. A new DNA (RNA) strand is synthesized by one or more
DNA (RNA) polymerase enzymes by means of a template consisting of a
pre-existing polynucleotide strand complementary to it in the sense
of Watson-Crick base pairing. During the course of this process, a
given polymerase may recognize its  $4$ distinct nucleobase ligands
in the template over one of the three different edges on
nucleobases, namely Watson-Crick, Hoogsten, and sugar edges
appearing respectively at base pairing, major groove and minor
groove sides of the double-helical structure of the nucleic acid.
Such an enzyme can form at most $2$ or $3$ H-bonds with its ligands
if the molecular recognition occurs over the Watson-Crick edges,
whereas no more than $1$ H-bond can be formed during the
protein-ligand interactions in the minor groove. However, polymerase
enzymes are not restricted to participate in their intermolecular
H-bonds with the same set of proton-acceptors and donors as opposed
to the spurious molecule $A$, which requires $n \approx 1.47$
partially covalent H-bonds to distinguish  $4$ ligands in our
scenario.

Besides this, proteins that are responsible for various biological
functions exerted inside a cell are made up from only $20$ of the
more than $700$ amino acids. There is one special aminoacyl-tRNA
synthetase enzyme (aaRS) for each of these $20$ amino acids (with
some exceptions) in bacteria. In eukaryotes, cytoplasmic aaRSs and
mitochondrial aaRSs are different from each other, but none of them
is generally able to be loaded by more than one particular amino
acid as well. These enzymes transfer their cognate amino acids onto
acceptor stems of their cognate tRNA molecules in the first step of
protein synthesis. To do so, a given aaRS should single out $1$,
$2$, $4$, or $6$ tRNA molecules from the $64$ possibilities
depending on to which amino acid it is assigned. These molecular
recognition processes are generally determined by only a couple of
nucleotides from among the first four nucleobase pairs at the
acceptor stem together with the preceding unpaired nucleobase at
position $73$ according to the conventional tRNA numbering system
\cite{RNACode_1993, RNACode_2001}. Additionally, aaRSs are divided
into two classes each of which contains enzymes specific for $10$ of
the amino acids and has a distinct activity mechanisms such that
class I enzymes dock onto the minor groove side of the acceptor
stem, whereas class II enzymes approach the helical oligonucleotides
from the major groove side \cite{RNACode_2001}. Hence, an aaRS is
capable to form a few intermolecular H-bonds to discriminate its
tRNA ligands against all the others. However, likewise the
polymerase enzymes, these synthetase enzymes also do not form their
intermolecular H-bonds over the same set of proton-acceptors and
donors. On the contrary, the spurious molecule $A$ proposed in our
scenario distinguishes $6$ ligands by only $n \approx 1.77$
partially covalent H-bonds associated with the same proton-acceptors
and donors residing its binding site.

\section{A discussion on the primordial synthetases of the pre-DNA
world}

aaRSs are the vital elements of protein synthesis as they provide a
physical basis for the connection between the receipt encoded by
nucleic acids and the product made by amino acids. Note that they
are also products of the protein synthesis itself. As a result of
this, evolutionary history of these enzymes and their ligands has
received a lot of attention in the literature. On the one hand, the
3-dimensional region of tRNA acceptor stems recognized by aaRSs is
suggested to constitute an operational RNA code, which may have been
a predecessor of the contemporary genetic code in RNA world
\cite{RNACode_1993}. On the other hand, origin of aaRSs is widely
dated back to before the split of three kingdoms of life known as
Bacteria, Archae, and Eucarya \cite{RNACode_2001}. Since no
structural or functional resemblance has been observed between the
two aaRS classes, but conserved active site domains in all members
of a given class have familiar structure and sequence motifs
\cite{RNACode_1993, RNACode_2001, RNACode_2010}, each class may have
evolved from a different ancestor.

Nonconservative domains of the modern aaRSs were probably added to
their ancestors later in their evolution and intertwined with the
emergence of the anticodon domain in tRNA ligands
\cite{RNACode_1993}. Hence, primordial synthetases of the pre-DNA
world may have been smaller proteins having less atoms in their
active sites. Also, it is trivial that they were likely to be
responsible for the discrimination of too many ligands when compared
to modern aaRSs, e.g., they may have not specialized for only one
amino acid and their nucleotide ligands may have not been limited to
tRNAs, but may have also included some cellular RNAs. In a sense,
progenitors of modern aaRSs may have been restricted in recognizing
their ligands to the same extent as the spurious molecule $A$ is
restricted in our scenario. On that account, these ancient enzymes
may have utilized the quantum entanglement shared in intermolecular
H-bonds in a way similar to that $A$ does to change its
configuration upon binding to an agonist ligand. That is, quantum
entanglement may have played a pivotal role in the emergence of
genetic machinery when the first enzymes were discriminating more
nucleotides using less H-bonds.
%

\section{Conclusions}

Using the tools of quantum information theory, we approached the
back and forth quantum tunneling events taking place between
chemically bonded atoms. We showed that tunneling of an electron
pair that defines a covalent bond cannot generate a useful quantum
entanglement because of the indistinguishability of the electrons.
However, if a covalent bond possesses a particular amount of ionic
character, then it can acquire a useful quantum entanglement as
well. This is especially expected in a covalent bond formed between
an electronegative atom and a H atom. Also, if such a covalent bond
participates in a classical H-bond by electrostatically interacting
with another electronegative atom, it gains much more ionic
character and so the amount of useful entanglement is enhanced.
Conversely, we demonstrated that tunneling of either the electrons
of the proton-acceptor atom or the proton of the H atom can generate
useful quantum entanglement in a H-bonded system, and the amount of
this entanglement increases with a decrease in the amount of ionic
character of the proton-donating covalent bond. Also, a significant
amount of entanglement can be shared between the atoms in such a
partially covalent H-bond even in the thermal equilibrium.

Could it be possible for biological organisms to utilize the useful
entanglement shared in partially covalent H-bonds? To explore this
question, we constructed a hypothetical molecular recognition
scenario in which a receptor protein or an enzyme recognizes its
ligands using $n$ intermolecular H-bonds involving proton or
electron delocalization. We restricted the biomolecule under
consideration to participate in each of these H-bonds with the same
set of proton-acceptors and donors residing on its binding site. We
found that $3^n - 1$ agonist ligands can be discriminated by the
biomolecule when the intermolecular entanglements are teleported
into its intramolecular H-bonds by means of the effect of its
conformation on both its intra- and intermolecular H-bonds.

Discrimination of $3^n - 1$ ligands under such a restricted
circumstance corresponds to an enormously high efficiency of the
usage of H-bonds as resource in molecular recognition. To reveal
this more precisely, we summarized the molecular recognition
processes involved in DNA replication and protein synthesis.
Although these processes are quite accurate and very fast, our
spurious biomolecule was found to be superior than the polymerase
and synthetase enzymes in the sense that it singles out more ligands
using less H-bond acceptors and donors in ligand recognition.
Finally, we considered the primordial ancestors of the contemporary
synthetases that may have lived in pre-DNA world. As these ancient
enzymes were probably discriminating more ligands using less
H-bonds, we argued that quantum entanglement shared in
synthetase-ligand H-bonds may have been somehow used as resource in
molecular recognition until the emergence of more complex enzymes in
the evolution of genetic machinery.

\newpage

\section*{Acknowledgments}

O.P and C.D. thank Tristan Farrow, Vlatko Vedral, and Keith Burnett
for many insightful discussions that motivated this paper. O.P. also
thanks Vlatko Vedral and members of his group in University of
Oxford for their hospitality during the earlier stages of this
research. Finally, O.P. and G.T. acknowledge financial support by
the Scientific and Technological Research Council of Turkey
(TUBITAK).

\appendix*

\section{Entanglement swapping protocol}

In what follows, we will construct a simple protocol for the
two-step state transformation given as
\begin{eqnarray}\begin{aligned} \label{Eq_EntSwap}
\ket{\varphi} \equiv
\big(\alpha_{+}\ket{10}&+\beta_{+}\ket{00}+\gamma_{+}\ket{01}\big)_{{\text{X}_N}{\text{X}_1}}
\otimes\Big(\frac{\ket{1}-\ket{0}}{\sqrt{2}}\Big)_{\text{X}_2}& \\
&\downarrow \text{Measurement on X$_N$ in the basis of $\{\lambda_1
|1\rangle + \lambda_2 |0\rangle, \lambda_2^* |1\rangle - \lambda_1^* |0\rangle\}$} &\\
&\downarrow \text{Local operations on joint X$_1$X$_2$ system depending on the measurement outcome} &\\
\big(\alpha_{+}\ket{10}&+\beta_{+}\ket{00}+\gamma_{+}\ket{01}\big)_{{\text{X}_2}{\text{X}_1}}
\otimes \ket{\psi}_{\text{X}_N} \equiv \ket{\psi} ,&
\end{aligned}\end{eqnarray}
where  $|\alpha_+|^2 + |\delta_+|^2 + |\beta_+|^2 = 1$,
$|\lambda_1|^2 + |\lambda_2|^2 = 1$, $N = \{B, C\}$,
$\ket{\psi}_{\text{X}_B} = \lambda_1 |1\rangle + \lambda_2
|0\rangle$, and $\ket{\psi}_{\text{X}_B} = \lambda_1^* |1\rangle -
\lambda_1^* |0\rangle$.

Assume without any loss of generality that $\beta_+ = 0$ and
$\lambda_1 = \lambda_2 = 1/\sqrt{2}$. The former assumption
maximizes the entanglement shared in the initial intermolecular
H-bond by minimizing the ionic character of the proton donating
X$_N-$H covalent bond during the interaction with proton-acceptor
X$_1$. On the other hand, the latter assumption attributes the same
amount of ionic character to both X$_B-$H and X$_C-$H covalent bonds
in the absence of any interaction with X$_1$.

Note that the computational basis states can be written as
\begin{equation}
\ket{0}=\frac{\ket{+}+\ket{-}}{\sqrt{2}}, \quad
\ket{1}=\frac{\ket{+}-\ket{-}}{\sqrt{2}},
\end{equation}
where $\ket{\pm}=\frac{1}{\sqrt{2}}(\ket{0}\pm\ket{1})$. Before
making a measurement on X$_N$ in the basis of $\{\ket{\pm}\}$, let
us rewrite the initial state in this basis as
\begin{eqnarray}
\ket{\varphi}
&=&\frac{1}{\sqrt{2}}\big(\alpha_{+}\ket{101}-\alpha_{+}\ket{100}
+\gamma_{+}\ket{011}-\gamma_{+}\ket{010}\big)_{\text{X}_N{\text{X}_1}{\text{X}_2}} \nonumber \\
&=&\frac{1}{\sqrt{2}}\Big(\alpha_{+}(\frac{\ket{+}-\ket{-}}{\sqrt{2}})\ket{01}-\alpha_{+}(\frac{\ket{+}-\ket{-}}{\sqrt{2}})\ket{00}
+\gamma_{+}(\frac{\ket{+}+\ket{-}}{\sqrt{2}})\ket{11} - \gamma_{+}(\frac{\ket{+}+\ket{-}}{\sqrt{2}})\ket{10}\Big)_{\text{X}_N{\text{X}_1}{\text{X}_2}} \nonumber \\
&=&\frac{1}{2}\Big(\ket{+}\big[\alpha_{+}\ket{01}-\alpha_{+}\ket{00}+\gamma_{+}\ket{11}-\gamma_{+}\ket{10}\big]
-
\ket{-}\big[\alpha_{+}\ket{01}-\alpha_{+}\ket{00}-\gamma_{+}\ket{11}+\gamma_{+}\ket{10}\big]\Big)_{\text{X}_N{\text{X}_1}{\text{X}_2}}
.
\end{eqnarray}

It is then straightforward to show that the post-measurement state
of X$_1$X$_2$ joint system becomes
\begin{eqnarray} \label{Eq_PostMeasuredX1X2}
\ket{\psi}_{{\text{X}_1}{\text{X}_2}}
&=&\frac{1}{\sqrt{2}}\Big(\alpha_{+}\ket{01}-\alpha_{+}\ket{00}+\gamma_{+}\ket{11}-\gamma_{+}\ket{10}\Big)_{{\text{X}_1}{\text{X}_2}}
,
\end{eqnarray}
if $\ket{\psi}_{\text{X}_N}$ is found to be $\ket{+}$ by the
measurement, while it equals to
\begin{eqnarray}
\ket{\psi}_{{\text{X}_1}{\text{X}_2}}
&=&\frac{1}{\sqrt{2}}\Big(\alpha_{+}\ket{01}-\alpha_{+}\ket{00}-\gamma_{+}\ket{11}+\gamma_{+}\ket{10}\Big)_{{\text{X}_1}{\text{X}_2}}
,
\end{eqnarray}
otherwise.

The second step of the protocol given in (\ref{Eq_EntSwap}) starts
with a conditional operation applied only on X$_2$ depending on the
measurement outcome: if $\ket{\psi}_{\text{X}_N}$ is measured as
$\ket{-}$, X$_2$ is subjected to a Pauli $Z$ operation, a unitary
transformation that equals to $\ket{0}\bra{0}-\ket{1}\bra{1}$. The
conditional operation defined in this way brings the state of
X$_1$X$_2$ joint system into (\ref{Eq_PostMeasuredX1X2}) in either
case, and is followed by a joint unitary operation on X$_1$X$_2$
which implies the state transformations to be
\begin{eqnarray} \begin{aligned} \label{Eq_EntSwap_UX1X2}
\frac{1}{\sqrt{2}}(\ket{01}-\ket{00})_{\text{X}_1\text{X}_2}
&\rightarrow \ket{01}_{\text{X}_1\text{X}_2} , \quad
\frac{1}{\sqrt{2}}(\ket{11}+\ket{10})_{\text{X}_1\text{X}_2}
\rightarrow \ket{11}_{\text{X}_1\text{X}_2} , \\
\frac{1}{\sqrt{2}}(\ket{11}-\ket{10})_{\text{X}_1\text{X}_2}
&\rightarrow \ket{10}_{\text{X}_1\text{X}_2} , \quad
\frac{1}{\sqrt{2}}(\ket{01}+\ket{00})_{\text{X}_1\text{X}_2}
\rightarrow \ket{00}_{\text{X}_1\text{X}_2} .
\end{aligned} \end{eqnarray}

After these two subsequent operations, the X$_1$X$_2$ joint system
ends up in the target state given by
\begin{eqnarray}
\ket{\psi}_{{\text{X}_1}{\text{X}_2}}
&=&\big(\alpha_{+}\ket{01}+\gamma_{+}\ket{10}\big)_{{\text{X}_1}{\text{X}_2}}
.
\end{eqnarray}

As Pauli $Z$ operation implies the state transformations $\ket{-}
\leftrightarrow \ket{+}$, its application on X$_2$ in the beginning
of the second step of the protocol physically corresponds to a
switch between two distinguishable states of X$_2-$H covalent bond
each of which possesses $50$ percent ionic character. Besides this,
the two state transformations given in the first line of
(\ref{Eq_EntSwap_UX1X2}) physically correspond to disappearance of
the ionic character of X$_2-$H covalent bond. Conversely, the two
state transformations given in the next line describe a process in
which X$_2-$H covalent bond breaks down and becomes a completely
ionic bond. Note that none of the input states of the joint unitary
operation defined in (\ref{Eq_EntSwap_UX1X2}) is entangled, but
X$_2-$H covalent bond possesses $50$ percent ionic character in each
of them. Also note that the initial state of X$_2$ in
(\ref{Eq_EntSwap}) is ionic to the same extent. Hence, all the
quantum operations involved in the present entanglement swapping
protocol, expect the initial measurement on X$_N$, are somehow
related to the amount of ionic character of the X$_2-$H covalent
bond, which may take the values of $0$, $50$, or $100$ percent. We
propose that, likewise the measurements discussed in Section
\ref{Sec_Measurments}, these operations can be also regarded as an
influence of the conformational dynamics on the X$_2-$H covalent
bond.


\begin{thebibliography}{99}

\bibitem{1989_PRA_Lloyd_IAndS} Lloyd, S, 1989. Use of mutual information
to decrease entropy: implications for the second law of
thermodynamics. \textit{Phys. Rev. A} \textbf{39}, 5378.

\bibitem{2012_PRL_SagawaAndUeda_CorrInSTD} Sagawa, T. and Ueda, M., 2012.
Fluctuation theorem with information exchange: role of correlations
in stochastic thermodynamics. \textit{Phys. Rev. Lett.}
\textbf{109}, 180602.

\bibitem{2014_PRL_SagawaEtAll_Exp} Koski, J. V., Maisi, V. F., Sagawa, T.,
and Pekola, J. P., 2014. Experimental observation of the role of
mutual information in the nonequilibrium dynamics of a Maxwell
demon. \textit{Phys. Rev. Lett.} \textbf{113}, 030601.

\bibitem{2013_HorodeckiOppenheim_Thermomajorization} Horodecki, M. and Oppenheim,
J. (2013). Fundamental limitations for quantum and nanoscale
thermodynamics. \textit{Nat. Commun.} \textbf{4}, 2059.

\bibitem{2015_HorodeckiOppenheim_2ndLaws} Brand\~{a}o, F.G.S.L., Horodecki, M.,
Ng, N.H.Y., Oppenheim, J. and Wehner, S., 2015. The second laws of
quantum thermodynamics. \textit{Proc. Natl. Acad. Sci. U.S.A.}
\textbf{112}, 3275--3279.

\bibitem{2017_Wehner_2ndLaws} van der Meer, R., Ng, N.H.Y. and Wehner, S., 2017.
Smoothed generalized free energies for thermodynamics. \textit{Phys.
Rev. A} \textbf{96}, 062135.

\bibitem{2017_NComms_Winter_UniversalLawsofTD} Bera, M. N., Riera,
A., Lewenstein, M., and Winter, A., 2017. Generalized laws of
thermodynamics in the presence of correlations. \textit{Nat.
Commun.} \textbf{8}, 2180.

\bibitem{2008_PRE_Partovi} Partovi, M. H., 2008. Entanglement versus stosszahlansatz:
disappearance of the thermodynamic arrow in a highcorrelation
environment. \textit{Phys. Rev. E} \textbf{77}, 021110.

\bibitem{2010_PRE_JenningsAndRudolph} Jennings, D. and Rudolph, T., 2010.
Entanglement and the thermodynamic arrow of time. \textit{Phys. Rev.
E} \textbf{81}, 061130.

\bibitem{2017_arXiv_Lutz} Micadei, K., Peterson, J. P. S., Souza, A. M., Sarthour, R. S.,
Oliveira, I. S., Landi, G. T., Batalh\~{a}o, T. B., Serra, R. M.,
and Lutz, E., 2017. Reversing the thermodynamic arrow of time using
quantum correlations. arXiv:1711.03323v1 [quant-ph].

\bibitem{1960_Pauling} Pauling, L. 1960, \textit{The Nature
of the chemical bond}, 3rd eds. Cornell University Press, Ithaca,
NY.

\bibitem{2011_RevCovOFHB} Grabowski, S. J.
2011, What is the covalency of hydrogen bonding? \textit{Chem. Rev.}
\textbf{111}, 2597--2625.

\bibitem{2002_BlueShiftedHB} Li, X., Liu, L., and Schlegel, H. B.,
2002. On the physical origin of blue-shifted hydrogen bonds.
\textit{J. Am. Chem. Soc.} \textbf{124}, 9639--9647.

\bibitem{1999_CovOfHBInDNA} Fonseca Guerra, C., Bickelhaupt, F. M.,
Snijders, J. G., and Baerends, E. J., 1999. The nature of the
hydrogen bond in DNA base pairs: The role of charge transfer and
resonance assistance. \textit{Chem. Eur. J} \textbf{5}, 3581--3593.

\bibitem{2000_CovOfHBInDNA} Fonseca Guerra, C., Bickelhaupt, F. M.,
Snijders, J. G., and Baerends, E. J., 2000. Hydrogen bonding in DNA
base pairs: Reconciliation of theory and experiment. \textit{J. Am.
Chem. Soc.} \textbf{122}, 411--128.

\bibitem{2006_CovOfHBInDNA} van der Wijst, T., Fonseca Guerra, C.,
Swart, M., and Bickelhaupt, F. M., 2006. Performance of various
density functionals for the hydrogen bonds in DNA base pairs.
\textit{Chem. Phys. Lett.} \textbf{426}, 415--421.

\bibitem{LBHB_1998_MiniRev} Cleland, W. W., Frey, P. A., and Gerlt, J.
A., 1998. The low barrier hydrogen bond in enzymatic catalysis.
\textit{J. Biol. Chem.} \textbf{273}, 25529--25532.

\bibitem{LBHB_1998_SSHB} Schi{\o}tt, B., Iversen, B. B., Madsen, G. K. H.,
Larsen, F. K., and Bruice, T. C., 1998. On the electronic nature of
low-barrier hydrogen bonds in enzymatic reactions. \textit{Proc.
Natl. Acad. Sci. USA} \textbf{95}, 12799--12802.

\bibitem{LBHB_2004_No} Schutz, C. N. and Warshel, A., 2004. The low barrier
hydrogen bond (LBHB) proposal revisited: The case of the
Asp$\cdot\cdot\cdot$His pair in serine proteases. \textit{Proteins}
\textbf{55}, 711--723.

\bibitem{LBHB_2006_No} Warshel, A., Sharma, P. K., Kato, M., Xiang, Y.,
Liu, H., Olsson, M. H. M., 2006. Electrostatic basis for enzyme
catalysis. \textit{Chem. Rev.} \textbf{106}, 3210--3235.

\bibitem{LBHB_2012} Elias, M., Wellner, A., Goldin-Azulay, K., Chabriere, E.,
Vorholt, J. A., Erb, T. J., and Tawfik, D. S., 2012. The molecular
basis of phosphate discrimination in arsenate-rich environments.
\textit{Nature} \textbf{491}, 134--137.

\bibitem{LBHB_2013} Hosur, M. V.,  Chitra, R., Hegde, S., Choudhury, R. R.,
Das, A., and Hosur, R. V., 2013. Low-barrier hydrogen bonds in
proteins. \textit{Crystallogr. Rev.} \textbf{19}, 3--50.

\bibitem{1963_HBPT_Lowdin} L\"{o}wdin, P. O., 1963. Proton tunneling in DNA
and its biological implications. \textit{Rev. Mod. Phys.}
\textbf{35}, 724--732.

\bibitem{2005_Villani} Villani, G., 2005.
Theoretical investigation of hydrogen transfer mechanism in the
Adenine-Tymine base pair. \textit{Chem. Phys.} \textbf{316}, 1--8.

\bibitem{2006_Villani} Villani, G., 2006.
Theoretical investigation of hydrogen transfer mechanism in the
Guanine-Cytosine base pair. \textit{Chem. Phys.} \textbf{324},
438--446.

\bibitem{2010_Villani-1} Villani, G., 2010.
Theoretical investigation of the hydrogen atom transfer in the
Adenine-Tymine base pair and its coupling with electronic
rearrangement. Concerted $vs$. Stepwise mechanism. \textit{Phys.
Chem. Chem. Phys.} \textbf{12}, 2664--2669.

\bibitem{2010_Villani-2} Villani, G., 2010.
Theoretical investigation of the hydrogen atom transfer in the
Cytosine-Guanine base pair and its coupling with electronic
rearrangement. Concerted $vs$. stepwise mechanism. \textit{J. Phys.
Chem. B} \textbf{114}, 9653--9662.

\bibitem{2014_BrovaretsAndHovorun-1} Brovarets, O. O. and Hovorun, D.
M., 2014. Can tautomerization of the A$\cdot$T Watson-Crick base
pair via double proton transfer provoke point mutations during DNA
replication? A comprehensive QM and QTAIM analysis. \textit{J.
Biomol. Struct. Dyn.} \textbf{32}, 127--154.

\bibitem{2014_BrovaretsAndHovorun-2} Brovarets, O. O. and Hovorun, D.
M., 2014. Why the tautomerization of the G$\cdot$C Watson-Crick base
pair via the DPT does not cause point mutations during DNA
replication? QM and QTAIM comprehensive analysis. \textit{J. Biomol.
Struct. Dyn.} \textbf{32}, 1474--1499.

\bibitem{2015_BrovaretsAndHovorun} Brovarets, O. O. and Hovorun, D.
M., 2015. Proton tunneling in the A$\cdot$T Watson-Crick DNA base
pair: myth or reality? \textit{J. Biomol. Struct. Dyn.} \textbf{33},
2716--2720.

\bibitem{2015_Al-Khalili} Godbeer, A. D., Al-Khalili, J. S., and Stevenson, P.
D., 2015. Modelling proton tunnelling in the Adenine-Thymine base
pair. \textit{Phys. Chem. Chem. Phys.} \textbf{17}, 13034--13044.

\bibitem{2015_PTInGC} Turaeva, N. and Brown-Kennerly, V., 2015.
Marcus model of spontaneous point mutation in DNA. \textit{Chem.
Phys.} \textbf{461}, 106--110.

\bibitem{Wooters-1996} Bennett, C. H., DiVincenzo, D. P., Smolin,
J., and Wootters, W. K., 1996. Mixed-state entanglement and quantum
error correction. \textit{Phys. Rev. A.} \textbf{54}, 3824--3851.

\bibitem{1993_PRL_EntSwap} \.{Z}ukowski, M., Zeilinger, A. , Horne, M. A.,
and Ekert, A. K., 1993. ``Event-ready-detectors'' Bell experiment
via entanglement swapping. \textit{Phys. Rev. Lett.} \textbf{71},
4287--4290.

\bibitem{1998_PRA_EntSwap} Bose, S., Vedral, V., and Knight, P. L.,
1998. Multiparticle generalization of entanglement swapping.
\textit{Phys. Rev. A} \textbf{57}, 822--829.

\bibitem{1993_PRL_qTeleport} Bennett, C. H., Brassard, G., Cr\'epeau, C.,
Jozsa, R., Peres, A., and Wootters, W. K., 1993. Teleporting an
unknown quantum state via dual classical and Einstein-Podolsky-Rosen
channels. \textit{Phys. Rev. Lett.} \textbf{70}, 1895--1899.

\bibitem{RNACode_1993} Schimmel, P., Gieg{\'e}, R., Moras, D., and Yokoyama,
S., 1993. An operational RNA code for amino acids and possible
relationship to genetic code. \textit{Proc. Natl. Acad. Sci. USA}
\textbf{90}, 8763--8768.

\bibitem{RNACode_2001} Ribas de Pouplana, L. and Schimmel, P., 2001.
Operational RNA Code for Amino Acids in Relation to Genetic Code in
Evolution. \textit{J. Biol. Chem.} \textbf{276}, 6881--6884.

\bibitem{RNACode_2010} Shaul, S., Berel, D., Benjamini, Y., and Graur,
D., 2010. Revisiting the operational RNA code for amino acids:
Ensemble attributes and their implications. \textit{RNA}
\textbf{16}, 141--153.


\end{thebibliography}
\end{document}